\documentclass[preprint]{sig-alternate-05-2015} 

\usepackage[utf8]{inputenc}
\usepackage[T1]{fontenc}

\usepackage[english]{babel}

\usepackage{csquotes}

\usepackage{amsmath}

\usepackage{booktabs}
\usepackage{multirow}

\usepackage{numprint}
\usepackage[binary-units=true]{siunitx}

\newcommand\lstdots{\makebox[1em][c]{.\hfil.\hfil.}}

\usepackage{listings}
\usepackage{pifont}
\newcommand{\cmark}{\ding{51}}

\newcommand{\At}{\ensuremath{{\sf{{Attest}}}}}
\newcommand{\Hr}{\ensuremath{{\sf{{HashRenewal}}}}}

\newcommand{\AS}{\ensuremath{{\sf{AS}}}}
\newcommand{\MTS}{\ensuremath{{\sf{MTS}}}}
\newcommand{\MDS}{\ensuremath{{\sf{MDS}}}}
\newcommand{\SLS}{\ensuremath{{\sf{SLS}}}}
\newcommand{\NAW}{\ensuremath{{\sf{NAW}}}}

\newcommand{\MT}{\ensuremath{{\textit{MT}}}}
\newcommand{\MTroot}{\ensuremath{{\textit{Root}}}}
\newcommand{\MTauth}{\ensuremath{{\textit{AuthenticationPath}}}}

\makeatletter
\def\@copyrightspace{
	\@float{copyrightbox}[b]
		\begin{center}
			\setlength{\unitlength}{0.2pc}
			\begin{picture}(100, 6)
				\put(0,-0.95){\crnotice{\@toappear}}
			\end{picture}
		\end{center}
	\end@float
}
\makeatother

\toappear{* Please cite the conference version of this work published at ASIACCS'17 \cite{asiaccs2017}}

\begin{document}


\clubpenalty = 10000 
\widowpenalty = 10000

\title{MoPS: A Modular Protection Scheme for \linebreak Long-Term Storage}
\subtitle{(Full Version)*}

\numberofauthors{5}
\author{
	{Christian Weinert}\\
	\affaddr{TU Darmstadt, Germany} \\
	\email{christian.weinert@crisp-da.de}
\alignauthor
	{Denise Demirel}\\
	\affaddr{TU Darmstadt, Germany} \\
	\email{ddemirel@cdc.informatik.tu-darmstadt.de}
\alignauthor
	{Mart\'{i}n Vigil}\\
	\affaddr{UFSC, Brazil}\\
	\email{martin.vigil@ufsc.br}
\alignauthor
\and
\alignauthor
	{Matthias Geihs}\\
	\affaddr{TU Darmstadt, Germany} \email{mgeihs@cdc.informatik.tu-darmstadt.de}
\and
\alignauthor
	{Johannes Buchmann}\\
	\affaddr{TU Darmstadt, Germany} \email{buchmann@cdc.informatik.tu-darmstadt.de}
}

\maketitle

\begin{abstract}
Current trends in technology, such as cloud computing, allow outsourcing the storage, backup, and archiving of data.
This provides efficiency and flexibility, but also poses new risks for data security.
It in particular became crucial to develop protection schemes that ensure security even in the long-term, i.e.\ beyond the lifetime of keys, certificates, and cryptographic primitives.
However, all current solutions fail to provide optimal performance for different application scenarios.
Thus, in this work, we present MoPS, a modular protection scheme to ensure authenticity and integrity for data stored over long periods of time.
MoPS does not come with any requirements regarding the storage architecture and can therefore be used together with existing archiving or storage systems.
It supports a set of techniques which can be plugged together, combined, and migrated in order to create customized solutions that fulfill the requirements of different application scenarios in the best possible way.
As a proof of concept we implemented MoPS and provide performance measurements.
Furthermore, our implementation provides additional features, such as guidance for non-expert users and export functionalities for external verifiers.
\end{abstract}

\printccsdesc


\section{Introduction}
The development of solutions that allow to preserve important security goals such as authenticity and integrity even in the long run has become an important research direction.
During the last decades, the way documents are stored changed from secure offline media, e.g.\ hard disks, to company-wide document management systems.
Besides these interconnected systems, the increasing availability of reliable high-bandwidth Internet connections makes outsour\-cing of storage, backup, and archiving into the cloud increasingly attractive.
These technological trends allow for higher efficiency and flexibility, but also pose new risks for data security.

Classical protection schemes provide only a sufficient level of protection for the time interval these schemes are considered secure for the chosen parameters and keys.
Also, in practice, it must be considered that attackers might get access to keys, e.g.\ by stealing smartcards or hacking backup servers.

\paragraph{Related Work and Problem Description}
Up until now, various protection schemes that allow to archive data securely even in the long-term have been developed, i.e.\ the AdES family of schemes \cite{xades}, ERS \cite{ers1, ers2}, CIS \cite{cis}, CISS \cite{ipccc2014}, CN \cite{notarCN}, and AC \cite{notarAC}.
In \cite{cose2015}, the authors provide a rigorous summary of all existing protection schemes that provide long-term authenticity and integrity in archiving systems.
However, none of them provides an efficient solution for different scenarios which may appear in practice and the authors do not discuss solutions to this problem.

Assume, for instance, all electronic documents generated in a hospital should be stored such that their authenticity and integrity is protected.
If such large amounts of documents are generated and stored, ERS is currently the most efficient solution with respect to initializing and maintaining the protection.
However, when the record of a patient consisting of several files is opened and an efficient procedure for checking its authenticity and integrity is needed, CIS or CISS are the methods of choice.
Thus, a new scheme would be required which, at the same time, provides an efficient verification process for folders, like CIS or CISS, and an efficient process for generating and maintaining the data needed to protect the folders, like ERS.
For some scenarios it might be possible to identify exactly one approach after reviewing the state of the art.
However, besides the fact that an expert is needed who selects and implements the solution, another drawback of this approach is that the access pattern for documents might change over time.
While the patients are still alive, their medical records are opened and new files are added on a regular basis.
This makes CIS or CISS an interesting solution.
However, when a patient is deceased, the access pattern changes and a migration to a simple document protection, e.g.\ AdES, CN, or AC, would be more efficient.

\paragraph{Contribution}
The previously mentioned archiving schemes have the following shortcomings:
First, each scheme is only efficient for very specific access patterns.
This requires the existence of experienced users who are able to review the state of the art to choose the most suitable solution. 
Furthermore, in many cases the most efficient solution can only be gained by combining several techniques, thereby requiring even expert knowledge.
Second, there are scenarios in which access pattern change over time and no archiving scheme is available that supports migration.

To address these shortcomings we first analyzed the existing proposals for long-term archiving in order to extract reusable techniques and identified for which trust assumptions and access patterns they are most suitable.
The outcome is a toolbox of several techniques (presented in Section \ref{sec:modularization}), i.e.\ attestation techniques and data structures, that can be plugged together in order to instantiate customized protection schemes, thereby addressing many different application scenarios.
We focused on authenticity and integrity protection, as these two security goals are addressed with similar measures.
Performing a corresponding analysis for protection schemes providing long-term confidentiality is left for future work.

Furthermore, our toolbox comes with two features:
First, it allows combining data structures, thereby increasing the amount of possible protection schemes and correspondingly the amount of applications for which efficient solutions are available (presented in Section \ref{sec:combinations}).
Second, it allows migrating between different protections schemes, thereby addressing that application scenarios change over time (presented in Section \ref{sec:migration}).

Based on this toolbox we developed and implemented the first modular protection scheme for long-term storage, called MoPS.
Our implementation provides tools with graphical user interfaces for signing documents, protecting documents in the long-term, and verifying the protection of documents.
Furthermore, it comes with wizard-based guidance to support non-expert users when creating protection schemes and updating the protection of documents.
Note that this makes our solution a very important contribution for practical use since each user can set up and maintain customized storage solutions even without expert knowledge.
Finally, MoPS contains export functionalities for stored documents and their proofs of existence.

\paragraph{Structure}
This paper is organized as follows:
First, in Section \ref{sec:ltaip}, we explain how authenticity and integrity can be preserved in the long run.
Then, in Section \ref{sec:modularization}, we present our set of techniques, show how they can be combined in Section \ref{sec:combinations}, and explain how migrating between different configurations is possible in Section \ref{sec:migration}.
In Section \ref{sec:implementation} we provide details regarding our implementation and conclude with a summary and future work in Section \ref{sec:conclusion}.

\pagebreak

\section{Long-Term Authenticity and Integrity Protection}
\label{sec:ltaip}
To provide authenticity and integrity for electronic documents, digital signature schemes are used.
More precisely, when a document is stored, the document owner signs the document with its private signature key following the hash-and-sign paradigm\footnote{Using the hash-and-sign paradigm, the size of messages is reduced using a hash function and the hash value is signed.}.
This allows any third party retrieving the document to verify that it has been signed by the document owner (\emph{authenticity}) and that the document has not been modified (\emph{integrity}).
However, this solution does not provide authenticity and integrity in the long run, because signatures and hashes are only secure for a limited period of time.
One threat to the security of signature schemes is that a malicious party might get access to the private keys, e.g.\ by stealing smartcards.
Alternatively, an attacker can run a brute force attack by simply trying out all possible private keys, thereby identifying the used ones.
This attack also allows violating the security of hash functions, since they are only secure as long as an attacker is not able to find a collision, i.e.\ two documents which lead to the same hash value.
Furthermore, over time new attacks or technological progress may allow breaking signature schemes or hash functions even more efficiently.
In practice, the validity of signature keys is the limiting factor since they are only valid for some years.
This makes it very likely that the storage period of a document exceeds the time frame for which a signature remains secure.
In the following, we will denote the time period in which signature keys, signatures, or hash values are secure with the term \emph{validity period}.

To address that signatures come with a limited validity period and to provide authenticity and integrity even in the long run, the security of signatures is prolonged.
More precisely, before the validity period of a signature is about to end, a so-called \emph{proof of existence} for the signature and the document is generated.
A simple approach to create such a proof is to generate an attestation for the data by sending it to a trusted third party, e.g.\ a \emph{timestamping authority (TSA)}, which signs it together with the current time.
Thus, although an attacker might get access to the private signing key of the document owner enabling it to forge signatures, a verifier can distinguish whether a signature has been generated before or after the signature key became insecure.
Furthermore, the attacker cannot change signatures which have already been generated.
Therefore, the attestation prolongs the security of the signature and consequently the authenticity and integrity of the signed document.

However, also attestations come with a validity period determined by the generated signatures and hash values.
Thus, this procedure is performed repetitively. One example to accomplish this works as follows:
The initial attestation $a_0$ is generated by hashing the document together with its signature and signing the resulting hash value together with the current time.
We will refer to the procedure generating the initial attestation as \emph{initialization procedure}.
Then, when the security of the signature or the hash value is about to fade out, i.e.\ the validity period of $a_0$ is about to end, a \emph{renewal procedure} is performed during which an attestation $a_1$ for $a_0$ is generated.
In other words, the proof of existence consists of a chain $a_0, \dots, a_n$ of $n + 1$ attestations.
If the latest attestation $a_n$ is still secure, it proves that $a_{n - 1}$ existed at a point in time when the parameters used to generate $a_{n - 1}$ were still secure.
In the same way $a_{n - 1}$ proves the existence of $a_{n - 2}$, and so on, until $a_{0}$.
$a_0$ then proves the existence of the signature to a document, which in turn proves the authenticity and integrity of the document.

\section{Modularization of Techniques}
\label{sec:modularization}
Analyzing the state of the art in long-term archiving, there are mainly two criteria to distinguish different techniques.
First, the \emph{attestation techniques} used to generate the attestations.
Second, the \emph{data structures} specifying which information is encapsulated in the proofs of existence and which additional metadata must be stored.

\subsection{Attestation Techniques} \label{subsubsec:attestation_techniques}
The long-term archiving schemes proposed so far make use of three different techniques to generate attestations, where each technique comes with different security and trust assumptions.
Table \ref{tab:attestation_technique1} provides an overview of these assumptions as well as infrastructure requirements for the attestation techniques grouped by the corresponding issuers.
All approaches described below make use of cryptographic primitives which are only secure for a limited period of time for the chosen parameters.
Thus, all attestations generated with these techniques are only secure within a certain validity period.

\begin{table}
	\centering
	\caption{Assumptions and requirements for attestation techniques grouped by corresponding issuers.}
	\resizebox{\columnwidth}{!}{
    \begin{tabular}{@{}lccc@{}}
		\toprule
		Attestation Issuer							&	WVM		&	TSA		&	NA		\\
		
		\midrule

		\emph{Issuer Trust Assumptions}													\\
		\quad	WVM cannot be modified or deleted	&	\cmark	&			& 			\\
		\quad	Correct time included 				& 			&	\cmark	& 	\cmark	\\
		\quad	Input verified correctly			& 		 	&			&	\cmark	\\

		\addlinespace
		
		\emph{Security Assumptions}														\\
		\quad	Hash function security				& 	\cmark 	&	\cmark	&	\cmark	\\
		\quad	Signature scheme security			& 		 	&	\cmark	&	\cmark	\\

		\addlinespace
		
		\emph{Infrastructure Requirements}												\\
		\quad	Witnesses for issuing				&	\cmark	& 			&			\\
		\quad	Public key infrastructure			& 			&	\cmark	&	\cmark	\\
		
		\bottomrule
	\end{tabular}
	}
	\label{tab:attestation_technique1}
\end{table}

\subsubsection{WVM-Based Timestamps}
WVM-based timestamps \cite{bayer} are generated by publishing the input data, usually received in hashed form, on \emph{widely visible media} (WVM), e.g.\ bulletin boards or newspapers.
To verify such a timestamp, the verifier must check that the hash has been published at the claimed point in time.
The \emph{verification data} needed to perform this procedure depends on the instantiation of the WVM and is out of scope.

Using this attestation technique, it is necessary to trust in the correct functionality of the WVM.
More precisely, it must not be possible to delete or modify published data.
Furthermore, since only hashed data is published, it is assumed that the used hash function is secure for the chosen parameters within the validity period of the attestation.
When the security of the hash function is about to fade out, all data must be rehashed.
Furthermore, since these attestations inherit the time from the WVM, witnesses are needed testifying that the data has been published at the claimed point in time.

\subsubsection{Signature-Based Timestamps}
Signature-based timestamps are issued by a trusted third party called \emph{timestamping authority} (TSA).
More precisely, when receiving some input data, usually in hashed form, the TSA generates a timestamp by signing the data together with the current time.
To verify the correctness of this attestation, the verifier must check whether the signature generated by the TSA is correct and whether the certificate for the used signature key was still valid at the time the attestation got renewed.
It follows that the verification data should contain all data necessary to verify the certificate of the TSA, including revocation information collected at the time when the attestation got renewed.

Compared to WVM-based timestamps, this type of attestation can also be used for applications where no witnesses are available to testify the generation of attestations.
However, on the downside, the TSA is trusted to include the correct time, i.e.\ the time when the attestation was issued, in the timestamp.
Furthermore, a \emph{public key infrastructure} (PKI) is needed which binds the TSA to its public signature key by means of certificates.
Finally, it is necessary to trust in the security of the employed signature scheme and the hash function within the validity period of the attestation.

\subsubsection{Notarial Attestation}
\label{subsubsec:notarial_attestation}
Notarial attestations \cite{notarCN, notarAC} are issued by an instance that is able to verify certain properties of the received input before signing the data together with a date.
In the following we will refer to this instance as \emph{notarial authority} (NA).
Which properties are verified depends on the application and can, for instance, be the signature(s), and/or the certificate(s), and/or the attestation(s) of one or multiple signed documents.
This is also the main difference compared to signature- and WVM-based timestamps which are generated blindly for the received data.

More precisely, notarial attestations are generated as follows:
During the initialization procedure, the required properties of the input data are verified and, if valid, the data is signed together with the current time.
For most of the data structures presented in the next section, the attestations generated during the renewal procedure equal those of the initialization procedure.
The only exception is when using this technique together with the notarial attestation wrapper (see Section \ref{subsubsec:naw}).
In this case, after verification, the data is signed together with the time the proof of existence has been initialized.
During verification, the signatures of all notarial attestations contained in the proof of existence are verified and, if more than one attestation was generated, it is checked whether the certificates for the used signature keys were valid until the attestations got renewed.
Thus, like for signature-based timestamps, the verification data for attestations which got renewed should contain all data necessary to verify the certificate of the NA, including revocation information.

The advantage of notarial attestations is that since they allow to verify, for example, the correctness of document signatures before prolonging them and attestations before renewing them, it is more likely to detect failures.
However, this requires that significantly more and possibly sensitive data is sent to the NA and leads to a much higher computation and communication complexity.
Furthermore, since the NA issuing the notarial attestations is not restricted to include the current time in the attestations, it can, if malicious, backdate documents.
Thus, this instance must be trusted to verify the properties of the data received correctly and to include the appropriate time in the attestation.
Finally, like for signature-based timestamps, a PKI is needed which binds the NA to its public signature key.
Also, the employed signature scheme and hash function must be secure within the validity period of the attestation. 

\subsection{Data Structures}
\label{subsubsec:data_structures}
Attestation techniques are used to generate proofs of existence for signed documents.
These proofs allow a third party to verify that a given document has been generated by a specific data owner and that no unauthorized changes have been performed since.
To provide an efficient proof generation and verification process, several data structures were proposed.
However, whether a certain approach is indeed more efficient than another depends on the access pattern with which documents are stored and opened and, correspondingly, with which the proofs of existence for documents are generated and verified.
Furthermore, newer approaches propose improvements from which many schemes would benefit, but they were not retrospectively adapted.
Thus, we analyzed existing long-term archiving schemes, identified the individual data structures, and generalized and improved them.

In the following we first provide a generic description of the procedures used to generate and verify proofs of existence.
Then, we describe all identified data structures, \emph{attestation sequence}, \emph{Merkle tree sequence}, \emph{multiple documents sequence}, \emph{skip-list sequence}, and \emph{notarial attestation wrapper}, compare them with related work, and highlight for which access pattern which data structure is most suitable.
Table \ref{tab:data_structure1} summarizes the different data structures and the access patterns for which they are the most efficient solution.
We assume that choosing a data structure not only determines the way \emph{resources} (i.e.\ signed documents, attestations, and verification data) are stored, but also how the stored resources are hashed during the initialization, renewal, and verification procedures.

\begin{table}
	\caption{Overview of the different data structures and the access patterns they are most suitable for.}
	\resizebox{\columnwidth}{!}{\begin{tabular}{lccccc}
			\toprule
			Data Structure 					&	AS		&	MTS		&	MDS		&	SLS		&	NAW		\\
			
			\midrule
			
			\emph{Retrieval}							\\
			\quad	Single documents 		&	\cmark	&			&			&	\cmark	&	\cmark	\\
			\quad	Ranges of documents 	&			&			&	\cmark	&	\cmark				\\
			\quad	All documents 			&			&	\cmark 										\\

			\addlinespace

			\emph{Storage}																				\\
			\quad	Few documents 			&	\cmark	&			&			&			&	\cmark	\\
			\quad	Sets of documents		&			&	\cmark										\\
			
			\quad	Documents added to		&			&			&	\multirow{2}{*}{\cmark}	&	\multirow{2}{*}{\cmark}		\\
            \quad	folders sequentially 																\\
            
            \bottomrule
		\end{tabular}
	}
\label{tab:data_structure1}
\end{table}

While the notarial attestation wrapper can only be used with notarial attestations, all attestation techniques can be applied for the remaining data structures.
While WVM-based and signature-based timestamps are generated \enquote{blind\-ly}, the notarial attestation allows to check certain properties of the data before creating the attestation.
However, note that in this case the NA needs access to all data required to perform the checks.

\subsubsection{Generation and Verification Procedures}\label{sec:genericattest}
Each data structure comes with mainly three procedures: the \emph{initialization procedure}, the \emph{renewal procedure}, and the \emph{verification procedure}.

\textbf{Initialization Procedure.}
During the initialization, an attestation on input data $d = \mathcal{D} || s$ consisting of document $\mathcal{D}$ and its signature $s$ is generated as described in Section \ref{subsubsec:attestation_techniques}.
In the following we refer with $\At$ to a function that generates a WVM-based timestamp, a signature-based timestamp, or a notarial attestation by interacting with a third party.
This function is used to generate an initial attestation $a_0$ for the hash of input data $d$ and the current time $t_0$.
The data stored for verification is attestation $a_0$ and hash function $H_0$.

\textbf{Renewal Procedure.}
To prolong the security of the proof of existence, a new attestation $a_{n}$ is generated.
Here two cases are distinguished:
If the validity period of $a_{n - 1}$ is about to end, the \emph{attestation renewal procedure} is triggered and a new attestation is generated similarly to the process described for the initialization procedure.
If the security of the hash function $H_{n - 1}$ is about to fade out, the \emph{hash renewal procedure} is triggered.
Here the document and, depending on the data structure, also some additional data is rehashed using a new hash function $H_n$ and afterwards a new attestation for this hash value is generated.
Input to both procedures are at least the old attestation $a_{n - 1}$, up-to-date verification data $v_{n - 1}$, the time $t_n$ when the new attestation is generated, and additional data that depends on the concrete data structure.

\textbf{Verification.}
To verify the proof of existence for a signed document, first, all data that got attested by calling $\At$ is recomputed using the signed document, the hash functions, the verification data, and additional data that depends on the data structure. 
Then, the correctness of each attestation is verified by using the recomputed attested data and the verification data.
Finally, it is verified that each attestation was renewed before its validity period ended.
Since this procedure is equal for all data structures, it is not further detailed in the following descriptions.

\subsubsection{Attestation Sequence (AS)}
Using the data structure \emph{attestation sequence} (\AS), a proof of existence for a signed document $d$ is generated by creating a sequence of attestations $a_0, \dots, a_n$.
This process is basically as described in Section \ref{sec:genericattest} and a detailed description in form of pseudocode is given in Listing \ref{lst:as}.

\begin{lstlisting}[
	frame=single,
	caption={Attestation Sequence Procedures},
	label=lst:as
]
Initialization(InputData $d$, Time $t_0$, HashFunction $H_0$)
	$h_0$ = $H_0(d)$
	$a_0$ = Attest($H_0 || h_0, t_0$)
	Store $a_0$ and $H_0$ for verification

Renewal(InputData $d$, Attestations $a_0, \lstdots, a_{n - 1}$, VerificationData
	$v_0, \lstdots, v_{n - 1}$, Time $t_n$, HashFunctions $H_{n - 1}, H_n$)
	If $H_{n} = H_{n - 1}$
		$h_n$ = AttestationRenewal($a_{n - 1}$, $v_{n - 1}$, $H_{n}$)
	Else
		$h_n$ = HashRenewal(d, $a_0, \lstdots, a_{n - 1}$, $v_0, \lstdots, v_{n - 1}$, $H_{n}$)
	$a_{n}$ = Attest($H_n || h_n, t_n$)
	Store $a_n$, $v_{n - 1}$, and $H_n$ for verification

AttestationRenewal($a_{n - 1}$, $v_{n - 1}$, $H_{n}$)
	Return $H_n(a_{n - 1} || v_{n - 1})$

HashRenewal($d$, $a_0, \lstdots, a_{n - 1}$, $v_0, \lstdots, v_{n - 1}$, $H_{n}$)
	Return $H_n(d || a_0 || v_0 || \lstdots || a_{n - 1} || v_{n - 1})$
\end{lstlisting}

\label{sec:ades}
A similar data structure is used by the \emph{AdES family} that even comes with ETSI standards (e.g.\ ETSI TS 101 903 \cite{xades}).
However, an important difference between their solution and our component is that we distinguish whether only the validity period of the attestation or also of the hash function is about to end.
In the first case it is not necessary to choose a new hash function and to rehash all existing sequence elements when creating new attestations.
This leads to a much more efficient solution when generating and verifying the proof of existence.

\textbf{Pros and Cons.} The advantage of $\AS$ is that there is no need to store additional information specific to the data structure and there is also no need to hash additional information.
Thus, when single documents are protected, this data structure is the most efficient approach regarding computation and space complexity.
For a rigorous performance evaluation and a comparison between this approach and the other approaches see \cite{cose2015}\footnote{Although $\AS$ is an improved version of AdES, the main observations still apply.}.
The performance evaluations confirmed that if a huge amount of data must be protected, e.g.\ when storing or archiving medical records, a huge amount of attestations must be generated and updated.
It follows that this is only a suitable approach if proofs of existence are generated for very few documents and the proofs are verified individually.

\subsubsection{Merkle Tree Sequence (MTS)}
Using the data structure \emph{Merkle tree sequence} (\MTS), a proof of existence for a set of signed documents is generated.
More precisely, instead of using a hash function to hash a single document, a Merkle tree \cite{merkle} is used to generate a single hash value for a set of documents.
Then, the attestation $a_0$ is generated for the computed Merkle tree root $r_0$.
Note that when the hash function used to generate this Merkle tree is about to become insecure, also the security of the Merkle tree is about to fade out.
Thus, when $\Hr$ is triggered, a new Merkle tree for the content, i.e.\ of the previous Merkle tree, is computed and a new attestation is generated for its root.
The signed documents are stored together with the authentication paths of their leaves in the different Merkle trees to prevent that each Merkle tree must be completely recreated to verify single documents.
Furthermore, this still allows one to maintain a separate proof of existence for each signed document.
The pseudocode of this process can be found in Listing~\ref{lst:mts}.

\label{sec:ers}
A similar data structure is used by \emph{ERS} and its XML version \emph{XML\-ERS} which were standardized and can be found in the RFC standards RFC 4998 \cite{ers1} and RFC 6283 \cite{ers2}, respectively.
Their construction is different from our component with respect to the data used to compute the Merkle tree leaves when the hash function is renewed during the hash renewal procedure:
In our construction, we put only the documents and their authentication paths in the leaves.
The previous attestations and their verification data are rehashed and attested together with the tree root.
In ERS, also all attestations and their verification data are added to each leaf.
Since this data is equal for all documents, this leads to hashing redundant information, making our component more efficient with respect to the hash renewal procedure.

\textbf{Pros and Cons.}
Using our data structure $\MTS$, a proof of existence can be generated for a set of signed documents, thereby making this technique a good solution when huge amounts of documents need to be protected, since it speeds up the initialization and renewal procedure.
For an efficiency analysis of the usage of Merkle trees in archiving systems and a comparison with alternative techniques, see \cite{hpcc2014} and \cite{cose2015}.
The performance evaluation also showed that for use cases where only few data is protected, $\MTS$ is less efficient with respect to the renewal procedure and the space consumption since authentication paths must be generated and stored\footnote{Although $\MTS$ is an improved version of ERS, the main observations still apply.}.
Furthermore, this approach only provides an efficient solution if sets of documents are stored, but not for use cases where multiple documents of different sets are opened and verified.

\begin{lstlisting}[
	frame=single,
	caption={Merkle Tree Sequence Procedures},
	label=lst:mts
]
Initialization(InputData $d_0, \lstdots, d_{m - 1}$, Time $t_0$, HashFunction $H_0$)
	$\MT_0$ = ComputeMerkleTree($d_0, \lstdots, d_{m - 1}$) using $H_0$
	$r_0$ = $\MT_0.\MTroot$
	$a_0$ = Attest($H_0 || r_0, t_0$)
	Store $a_0$ and $H_0$ for verification
	For $i = 0, \lstdots, m - 1$
		$p_{i, 0}$ = $\MT_0.\MTauth_i$
		Store $p_{i, 0}$ with $d_i$ for recomputing $r_0$

Renewal(InputData $d_0, \lstdots, d_{m - 1}$, Attestations $a_0, \lstdots, a_{n - 1}$,
	VerificationData $v_0, \lstdots, v_{n - 1}$, AuthenticationPaths $p_{0, 0}$,
	$\lstdots, p_{m - 1, k - 1}$, Time $t_n$, HashFunctions $H_{n - 1}$, $H_{n}$)
	If $H_n$ = $H_{n - 1}$
		AttestationRenewal($a_{n - 1}$, $v_{n - 1}$, $t_n$, $H_{n}$)
	Else
		HashRenewal($d_0, \lstdots, d_{m-1}$, $a_0, \lstdots, a_{n - 1}$, $v_0, \lstdots, v_{n - 1}$,  
			$p_{0, 0}, \lstdots, p_{m - 1, k - 1}$, $t_n$, $H_{n}$)

AttestationRenewal($a_{n - 1}$, $v_{n - 1}$, $t_n$, $H_{n}$)
	$h_n$ = $H_n(a_{n - 1} || v_{n - 1})$
	$a_{n}$ = Attest($H_n || h_n, t_n$)
	Store $a_n$, $v_{n - 1}$, and $H_n$ for verification

HashRenewal($d_0, \lstdots, d_{m - 1}$, $a_0, \lstdots, a_{n - 1}$, $v_0, \lstdots, v_{n - 1}$,
	$p_{0, 0}, \lstdots, p_{m - 1, k - 1}$, $t_n$, $H_{n}$)
	$\MT_k$ = ComputeMerkleTree($d_0 || p_{0, 0} || \lstdots || p_{0, k - 1}, \lstdots,$
		$d_{m - 1} || p_{m - 1, 0} || \lstdots || p_{m - 1, k - 1}$) using $H_n$
	$r_k$ = $\MT_k.\MTroot$
	$h_n$ = $H_n(a_0 || v_0 || \lstdots || a_{n - 1} || v_{n - 1})$
	$a_n$ = Attest($H_n || r_{k} || h_n, t_n$)
	Store $a_n$, $v_{n - 1}$, and $H_n$ for verification
	For $i = 0, \ldots, m - 1$
		$p_{i, k}$ = $\MT_k.\MTauth_i$
		Store $p_{i, k}$ with $d_i$ for recomputing $r_k$
\end{lstlisting}

\subsubsection{Multiple Documents Sequence (MDS)}
Using the \emph{multiple documents sequence} (\MDS), a proof of existence for a batch of signed documents is generated where the documents are added subsequently to the storage system or archive.
After a proof of existence for the initial signed document has been generated, each time a new document is added to the batch, it is appended to the proof and an attestation for both, the new document and the proof, is generated.
Like for $\MTS$, the rehashing in $\Hr$ is done using Merkle trees to prevent that a verifier needs access to all signed documents, attestations, and their verification data contained in the chain in order to verify $a_n$.
For simplicity we assume that the time intervals in which documents are added are smaller than the validity periods of attestations.
This is a reasonable assumption, for instance, when using signature-based timestamps to securely store medical records.
Due to the employed signature schemes, the validity period of attestations holds at least two years while most patients consult their doctors multiple times a year.
However, if the security of the proof of existence must be prolonged without adding a new document, this can be performed by using the algorithms described for $\AS$. For the pseudocode see Listing \ref{lst:mds}.

\begin{lstlisting}[
	frame=single,
	caption={Multiple Documents Sequence Procedures (for $0 < j < n - 1$, $j$ denotes the penultimate iteration $\Hr$ was called)},
	label=lst:mds
]
Initialization(InputData $d_0$, Time $t_0$, HashFunction $H_0$)
	$h_0$ = $H_0(d_0)$
	$a_0$ = Attest($H_0 || h_0$, $t_0$)
	Store $a_0$ and $H_0$ for verification

Renewal(InputData $d_0, \lstdots, d_{n}$, Attestations $a_0, \lstdots, a_{n - 1}$,
	VerificationData $v_0, \lstdots, v_{n - 1}$, AuthenticationPaths $p_{0, 0}, \lstdots,$
	$p_{j, k - 1}$, Time $t_n$, HashFunctions $H_{n - 1}$, $H_{n}$)
	If $H_{n} = H_{n - 1}$
		AttestationRenewal($d_n, a_{n - 1}$, $v_{n - 1}$, $t_n$, $H_n$)
	Else
		HashRenewal($d_0, \lstdots, d_{n}$, $a_0, \lstdots, a_{n - 1}$, $v_0, \lstdots, v_{n - 1}$, 
			$p_{0, 0}, \lstdots, p_{j, k - 1}$, $t_n$, $H_{n}$)

AttestationRenewal($d_n$, $a_{n - 1}$, $v_{n - 1}$, $t_n$, $H_{n}$)
	$h_n$ = $H_n(H_n(d_n) || a_{n - 1} || v_{n - 1})$
	$a_n$ = Attest$(H_n || h_n, t_n)$
	Store $a_n$, $v_{n - 1}$, and $H_n$ for verification

HashRenewal($d_0, \lstdots, d_{n}$, $a_0, \lstdots, a_{n - 1}$, $v_0, \lstdots, v_{n - 1}$,
	$p_{0, 0}, \lstdots, p_{j, k - 1}$, $t_n$, $H_{n}$)
	$\MT_k$ = ComputeMerkleTree($d_0 || a_0 || v_0 || p_{0, 0} || \lstdots || p_{0, k - 1},$
		$\lstdots, d_{j} || a_{j} || v_{j} || p_{j, k - 1}, \lstdots, d_{n - 1} || a_{n - 1} || v_{n - 1}$) using $H_n$
	$r_k$ = $\MT_k.\MTroot$
	$h_n$ = $H_n(d_n)$
	$a_n$ = Attest($H_n || r_k || h_n, t_n)$
	Store $a_n$, $v_{n - 1}$, and $H_n$ for verification
	For $i = 0, \lstdots, n - 1$
		$p_{i, k}$ = $\MT_k.\MTauth_i$
		Store $p_{i, k}$ with $d_i$ for recomputing $r_k$
\end{lstlisting}

\label{sec:cis}
Also \emph{CIS} \cite{cis} allows generating one proof of existence for a batch of sequentially archived documents.
However, this scheme uses WVM-based timestamps and a data structure which looks like an unbalanced hash tree.
There are two significant differences between both approaches:
First, our data structure also provides an attestation renewal procedure which is needed for signature-based timestamps and which is not provided by CIS.
Second, during the hash renewal procedure we rehash the data contained in the sequence using a Merkle tree.
Thus, if verifiers want to verify a range of documents, they need only access to the documents that should be verified and the data added after the oldest document of the range.
Using CIS, access to all documents, attestations, and verification data is needed in order to verify a single document.

\textbf{Pros and Cons.}
Our data structure $\MDS$ allows generating one proof of existence for a batch of sequentially stored or ar\-chived documents.
In contrast to the data structure \MTS, the documents are not protected as a single set, but added sequentially to a common proof of existence.
This is an interesting technique for documents which have a content-related dependency, e.g.\ when archiving folders or records.
In this case, opening a folder containing several documents requires to verify only one chain of attestations where the number of attestations equals the number of documents protected.
However, if only single documents are opened, this approach is less efficient than the data structure \AS, see \cite{cose2015}\footnote{Although $\MDS$ is an improved version of CIS, the main observations still apply.}:
Each time a document is added, a new attestation is appended to the chain.
This leads to a much longer chain compared to the chain generated with the data structure $\AS$ where attestations are only appended when the validity period of old attestations is about to end.

\subsubsection{Skip-List Sequence (SLS)}
The data structure \emph{skip-list sequence} (\SLS) allows an efficient archiving solution for both access patterns, verifying ranges of documents and verifying single documents.
The drawback of $\MDS$ is that a new attestation is generated each time a document is added.
It follows that when retrieving a single document, several attestations with overlapping validity periods must be verified.
A possible solution to avoid this are append-only skip-lists \cite{skiplist} which maintain linked lists on multiple levels called \emph{parallel hash chains} allowing one to efficiently traverse lists of elements.
This allows at the same time generating one proof of existence for a sequence of documents and reducing the amount of attestations which must to be checked during the verification procedure.
The archiving scheme \emph{CISS} \cite{ipccc2014} provides such a skip-list-based solution.
Since MoPS uses this scheme without any modification and the individual processes are very complex, we refer for details to the original work.

\textbf{Pros and Cons.}
\SLS, like \MDS, allows one to retrieve and verify a batch of documents with a single attestation sequence and in addition it also allows one to retrieve single documents almost as efficient as when using the data structure \AS.
However, this comes at the cost of additional complexity regarding time when appending new elements (due to the need of creating the links) and storage space (due to the need of storing the links).
A rigorous performance analysis of this approach and a comparison with other data structures can  be found in \cite{ipccc2014}.
The performance analysis shows that using this data structure is only recommended when it is known in advance that in the given scenario there is a high probability that the validity periods of attestations will overlap.
In this case the benefit of skipping attestations during verification is greater than the cost of the additional complexity.

\subsubsection{Notarial Attestation Wrapper (NAW)}
\label{subsubsec:naw}
In contrast to all data structures described so far, the data structure \emph{notarial attestation wrapper} (\NAW) is a special case.
The attestation renewal procedure is performed by an NA which checks the correctness of the old attestation and if it is correct replaces it with a new attestation.
Thus, each proof of existence consists only of one single attestation.
During initialization, it must be clarified which property the NA is expected to verify.
The NA can, for instance, verify that the hash function $H_0$ and the certificate $c$ claimed to be used by the document owner are still secure.
Alternatively, the NA could also be asked to verify whether the signature key used to generate the signature $s$ to document $d$ indeed belongs to certificate $c$ and the signed document has been hashed using $H_0$.
During the renewal, the NA can check that the latest attestation $a_n$, i.e.\ its signature, is still valid and that the hash function $H_{n - 1}$ and, where applicable, also the new hash function $H_{n}$ is secure.
The pseudocode for the individual procedures can be found in Listing \ref{lst:naw}.
Note that except for the store and delete operations, it presents the view of the NA since the client only sends and receives data.

\pagebreak

\begin{lstlisting}[
	frame=single,
	caption={Notarial Attestation Wrapper Procedures},
	label=lst:naw
]
Initialization(HashValue $H_0(d)$, Certificate $c$, Time $t_0$,
	HashFunction $H_0$)
	If $c$ is valid at $t_0$ AND $H_0$ is secure at $t_0$
		$a_0$ = Attest($H_0 || H_0(d) || c, t_0$)
		Store $a_0$ and $H_0$ for verification
	Else
		Abort

Renewal(HashValues $H_0(d), \lstdots, H_n(d)$, Certificate $c$,
	Attestation $a_{n - 1}$, VerificationData $v_{n - 1}$, Times $t_0, t_n$,
	HashFunctions $H_0, \lstdots, H_n$)
	Verify $a_{n - 1}$ with $v_{n - 1}$
	If $a_{n - 1}$ is valid AND $H_n$ is secure at $t_n$
		If $H_n = H_{n - 1}$
			AttestationRenewal($H_0(d), \lstdots, H_{n - 1}(d)$, $c$, $a_{n - 1}$, $t_0$,
				$H_0, \lstdots, H_{n - 1}$)
		Else
			HashRenewal($H_0(d), \lstdots, H_n(d)$, $c$, $a_{n - 1}$, $t_0, t_n$,
				$H_0, \lstdots, H_n$)
	Else
		Abort

AttestationRenewal($H_0(d), \lstdots, H_{n - 1}(d)$, $c$, $a_{n - 1}$, $t_0$
	$H_0, \lstdots, H_{n - 1}$)
	$a_n$ = Attest($H_0 || H_0(d) || \lstdots || H_{n - 1} || H_{n - 1}(d) || c, t_0$)
	Store $a_n$ for verification
	Delete $a_{n - 1}$

HashRenewal($H_0(d), \lstdots, H_n(d)$, $c$, $a_{n - 1}$, $t_0, t_n$, $H_0, \lstdots, H_n$)
	If $H_{n - 1}$ is secure at $t_n$
		$a_n$ = Attest($H_0 || H_0(d) || \lstdots || H_{n} || H_{n}(d) || c, t_0$)
		Store $a_n$ and $H_n$ for verification
		Delete $a_{n - 1}$
	Else
		Abort
\end{lstlisting}

\emph{CN} \cite{notarCN} was the first approach that used this idea of notarial attestations.
However, it made use of two trusted third parties which are involved in the attestation generation process, namely TSAs and notaries.
Based on CN, an improved scheme called \emph{AC} \cite{notarAC} was developed.
This approach uses notarial attestations and a data structure which equals our data structure $\NAW$.

\textbf{Pros and Cons.}
Since the verifier only needs to verify one attestation, this approach is by far the most efficient data structure, see \cite{cose2015}.
However, the data structure $\NAW$ can only be used together with the attestation technique notarial attestation.
Furthermore, the NA generating the attestation should be a person with legal training who is licensed by the government to witness signatures on documents, such as a notary.
The proof of existence only contains the information at what time the proof of existence for a document has been initialized and who generated the latest attestation.
The information at what time and by which party the individual attestations have been renewed is lost.
Thus, all NAs that are involved in prolonging the security of the proof of existence must be trusted to verify the received data correctly and to include the correct time in the new attestation.
It follows that this data structure comes with much stronger trust assumptions compared to the data structures generating sequences of attestations.
Thus, before using this data structure for a use case, it must be evaluated whether these trust assumptions are feasible.

\section{Combination of Techniques}
\label{sec:combinations}
The data structures presented in the last section aim at providing efficient solutions for initializing, renewing and verifying proofs of existence.
On the one hand, $\AS$, $\MDS$, $\SLS$, and $\NAW$ ensure efficient verification procedures for different access patterns.
On the other hand, $\MTS$ provides efficient initialization and renewal procedures when proofs of existence for huge amounts of documents are generated at the same point in time.
Thus, in the following we discuss how for such large data sets $\MTS$ can be combined with the remaining data structures in order to gain an efficient solution with respect to all procedures.

First, we construct a combined data structure by cumulating the attestation requests of different data structures in one $\MTS$.
Then, we show that when large amounts of documents need to be archived, the data structures can be used to generate proofs of existence for Merkle tree roots instead of protecting single signed documents.

\subsection{Merkle Tree Sequences Combining Multiple Attestation Requests}
\label{subsec:mt_request}
$\MTS$ allows one to protect the integrity of a set of hash values with a single attestation sequence.
This technique cannot only be used for the hashes of signed documents, but also for the hash values generated during the initialization and renewal procedures of all data structures.

\textbf{Initialization Procedure.}
During the initialization, the hash values which need to be attested are computed by running the initialization or renewal procedures of $\AS$, $\MTS$, $\MDS$, $\SLS$, or $\NAW$.
Then, instead of calling $\At$ for the individual hash values, they are input for the initialization procedure of $\MTS$ which generates a Merkle tree and generates one attestation for the entire set of hash values.
However, note that if the leaves of the Merkle tree contain a hash generated with $\NAW$, then the Merkle tree root must be attested by an NA.
Furthermore, the same hash function should be used for all data structures combined in the Merkle tree in order to avoid executing the hash renewal procedure more often than inevitable.

\textbf{Renewal Procedure.}
Regarding the renewal of the attestation generated for the latest Merkle tree root, it must be considered which data structures belong to the hashes in the Merkle tree leaves.
$\MDS$ and $\SLS$ automatically renew attestations when new documents are added, while this is not the case for $\AS$, $\MTS$, and $\NAW$.
Thus, we must distinguish three cases.

In the first case, all hashes in the leaves are used for $\MDS$ or $\SLS$.
Then, the attestation renewal procedure of $\MDS$ or, respectively, $\SLS$ can be performed independently.
This is done as follows:
The Merkle tree root, the authentication path corresponding to the data structure, and the common attestations are appended to each data structure.
Then, the attestation renewal procedure of $\MDS$ or, respectively, $\SLS$ is executed as usual.
Cumulating the attestation requests when adding new documents to these data structures is not reasonable, except when new documents are added to all data structures at the same time.
When the hash renewal procedure instead of the attestation renewal procedure is triggered, all data contained in the respective data structures must be rehashed.
In MoPS, the hash renewal is performed for all data structures at the same time in order to be able to cumulate the attestation requests.
Since after the generation of the latest $\MTS$ new documents have been added to the individual sequences, the old $\MTS$ cannot simply be rehashed.
Thus, a new $\MTS$ is initialized by running the hash renewal procedures of the individual data structures and using the resulting hash values as input for the initialization procedure of the new $\MTS$.

\pagebreak

In the second case, all hashes in the leaves are used for $\AS$, $\MTS$, or $\NAW$.
Since these data structures do not renew attestations automatically, this is done by calling the attestation or hash renewal procedure provided by $\MTS$.
In case the hash renewal procedure is run and the Merkle tree is rehashed using a new hash function $H_n$, also the hashes in the leaves must be recomputed correspondingly.
Thus, the hash renewal procedure of $\AS$, $\MTS$, and, respectively, $\NAW$ is triggered and the data is rehashed using $H_n$.

In the third case, the hashes in the leaves belong to both, $\MDS$ or $\SLS$ and $\AS$, $\MTS$, or $\NAW$.
The data structures $\MDS$ and $\SLS$ renew their attestations as described for the first case.
The remaining data structures can use the attestation renewal technique as described for the second case.
When it is necessary to run the hash renewal procedure for all data structures, a new $\MTS$ is initialized as reasoned for the first case.

\textbf{Verification.}
Verifying a document protected by such a combined solution requires to additionally recompute the hash values added to $\MTS$.
Then, using the corresponding authentication paths, the roots of the Merkle trees are recomputed and the correctness of the attestations for the roots is verified by calling the verification procedure of $\MTS$.

\subsection{Data Structures Attesting Merkle Tree \\ Roots}
\label{sec:comb_trees}
Merkle trees can be combined with all data structures, i.e.\ $\AS$, $\MTS$, $\MDS$, $\SLS$, and $\NAW$.
However, since $\MTS$ results from combining the idea of Merkle trees with the data structure $\AS$, this combination will not be discussed explicitly.
Furthermore, the combination of $\MTS$ with $\MTS$ in order to cumulate attestation requests has already been discussed in the last section.

The data structures $\MDS$, $\SLS$, and $\NAW$ address the scenario where at a given point in time only one new document is attested.
Considering medical records as a possible use case, there are scenarios where this is not reasonable.
Assume, for instance, that a person got injured in an accident.
In this case, various tests must be conducted, e.g.\ a blood test, an x-ray scan, etc., and consequently a set of documents must be added to the record.
An efficient archiving solution for such a scenario can be provided by combining $\MDS$, $\SLS$, and $\NAW$ with Merkle trees.
Note that these data structures aim at providing the integrity of a given hash value which usually corresponds to a signed document.
This trivially allows protecting the root of a Merkle tree which contains hashes of multiple documents in the leaves instead.

Assume, for instance, a set of documents is added after an attestation $a_{n - 1}$ which has been generated using the initialization and renewal procedures of $\MDS$.
Then, if hash function $H_{n - 1}$ used to generate $a_{n - 1}$ is still secure, first, the Merkle tree root $r'_n$ is computed like in the initialization procedure of the data structure $\MTS$ using hash function $H_n = H_{n - 1}$.
Second, $a_{n}$ is computed as $a_n = \At(H_n || H_n(H_n(\\d_n) || a_{n - 1} || v_{n - 1}), t_n)$, where $d_n = r'_n$, $v_{n - 1}$ is the verification data needed to verify $a_{n - 1}$, and $t_n$ the time when $a_n$ is issued.
In case the security of hash function $H_{n - 1}$ is about to fade out, the Merkle tree root $r'_n$ is computed with a new hash function $H_n$.
Then, the same hash function is used to compute the Merkle tree root $r_k$ from all data contained in the sequence as described for $\MDS$.
Finally, the new attestation $a_n$ is computed as $a_n = \At(H_n || H_n(H_n(d_n) || r_k), t_n)$, where $d_n = r'_n$.
To be able to recreate the Merkle tree roots efficiently during verification, the authentication paths are stored with the documents.
The same way Merkle trees can be combined with the data structures $\SLS$ and $\NAW$.

The verification of a document requires to first recompute the Merkle tree root using the authentication path.
Then, the correctness of the attestation for the root is verified by running the verification procedure of the data structure $\MDS$, $\SLS$, or $\NAW$, respectively.

\section{Migration}
\label{sec:migration}
In general, there are two reasons why it is necessary to switch from one data structure to another.
The first one is that the requirements for the storage or archiving system changed.
Assume, for instance, the authenticity and integrity of a medical record is protected using $\MDS$ or $\SLS$.
Then, when the patient passed away, no new data will be added to the record, which is why the folder should be migrated to a simple data structure, such as $\AS$ or $\NAW$.
The second reason is that documents are transferred to another storage or archiving system which runs a different protection scheme.

How the migration procedure works depends on the target system configuration.
We first describe the migration procedure when the target system supports sequences, i.e.\ $\AS$, $\MTS$, $\MDS$, or $\SLS$.
Then, we suggest a construction for migrating to $\NAW$.

\subsection{Migration to Sequence-Based Data Structures}
\label{subsec:migration_ts}
Moving from an arbitrary scheme to a sequence-based data structure, i.e.\ $\AS$, $\MTS$, $\MDS$, or $\SLS$, requires migrating the proof of existence generated with the old data structure.
More precisely, assume a proof of existence has been generated and the latest attestation is $a_n$.
Then, we distinguish whether this is a proof protecting a single document, i.e.\ with $\AS$ or $\NAW$, or a set of documents, i.e.\ with $\MTS$, $\MDS$, or $\SLS$.
In the first case the proof is migrated by simply calling the renewal procedure of the target data structure for $a_n$.
This process automatically transfers the proof of existence in the new format and provides a time reference for the migration.
If the data structure protects multiple documents, first a Merkle tree is generated where each leaf contains one document and its proof of existence.
Then, the root of the Merkle tree is added to the new data structure.
Note that in both cases, when the hash renewal procedure of the new data structure is called, also the migrated data is rehashed.

If the target data structure allows to add documents, e.g.\ $\MDS$ or $\SLS$, the migrated data can also be added to an existing proof of existence.
This allows to merge folders or subsequently add older documents to existing folders.
In this case, signed documents and their proof of existence are hashed and added as new data to a proof of existence maintained by the target system.

The verification of a migrated document is performed in two steps.
First, the old proof of existence for the document is verified by running the verification procedure of the data structure used before the migration.
Then, the verification procedure of the target data structure is performed to verify that the data has been correctly protected after the migration.

\pagebreak

\subsection{Migration to NAW}
\label{subsec:migration_ac}
When migrating from a sequence-based data structure to $\NAW$, all documents together with their proofs of existence are sent to an NA.
The NA verifies the received data as specified by the target system and migrates the data.
Depending on the use case, it either returns a single notarial attestation for each document of the old data structure or it generates a Merkle tree with the documents as leaves and returns a single notarial attestation for the root of the tree (see Section \ref{sec:comb_trees}).
The new attestation(s) prove(s) the authenticity and integrity of the signed document(s).
However, note that the proof(s) of existence generated before the migration is/are deleted after verification.
Furthermore, the time $t_0$ contained in the notarial attestation does not refer to the time the data has been migrated, but to the date when the first attestation for the corresponding document or the set of documents was created.

\section{Implementation}
\label{sec:implementation}
In this section, we provide details regarding our implementation of MoPS.
It consists of a web application which provides the main user interface for a web service-based architecture and two additional desktop applications for creating signatures and verifying stored documents, respectively.

The section is organized as follows:
First we describe the features supported by our implementation in Section \ref{sec:requirements_workflow}.
In the following sections we present the three applications which together form MoPS: the signing application in Section \ref{sec:signing_app}, the web application in Section \ref{sec:web_app}, and the verification application in Section \ref{sec:verification_app}.
In Section \ref{sec:web_services}, the service architecture and the individual web services behind the web application are explained.
Finally, a short performance evaluation in Section \ref{subsec:performance} reports the computation time and the storage space consumption when using our implementation to protect documents.

\subsection{Features}
\label{sec:requirements_workflow}
Our current implementation covers the following attestation techniques, data structures, combinations, and migrations: 

Regarding attestation techniques, we support signature-based timestamps and notarial attestations.
WVM-based timestamps are not provided, because, first, to the best of our knowledge, there is currently no provider offering WVM-based timestamping services free of charge.
Second, we do not have access to WVM and therefore cannot create a realistic prototype service.
Notarial attestations in our implementation attest during initialization that the received certificate is valid and the employed hash function is secure.
When renewing an attestation, they attest that its signature is still valid and the employed hash function(s) is/are secure, just as described in Section \ref{subsubsec:naw}.
The supported cryptographic primitives for generating the attestations are the SHA-2 family of hash functions (SHA-256, SHA-384, SHA-512) and the RSA signature scheme where the length of the signature key depends on the chosen hash function (see Table \ref{tab:digest_methods_rsa_key_lengths} in Appendix \ref{subsubsec:ca} for details).

Regarding data structures, our implementation supports $\AS$, $\MTS$, $\MDS$, $\SLS$, and $\NAW$.
Furthermore, we implemented the unbalanced hash tree used by the archiving \linebreak scheme CIS for testing purposes.
We support the combination of $\MDS$ and $\SLS$ with Merkle trees exactly as described in Section \ref{sec:comb_trees}.
The combination of multiple attestation requests using $\MTS$ is left for future work.

Regarding migration, both of the migration procedures presented in Section \ref{sec:migration} are supported.
However, when migrating to $\NAW$, always a single notarial attestation is returned for each protected document.
The reason for this restriction is that, as explained previously, our notarial attestations attest the validity of single certificates and therefore cannot be used to attest an unsigned Merkle tree root.

Besides these basic MoPS features, our implementation also provides import and export functionalities which allow transferring protected documents between different instances of the system without losing protection.

\subsection{MoPS Apps}
\begin{figure}
\centering

\includegraphics[width = \columnwidth]{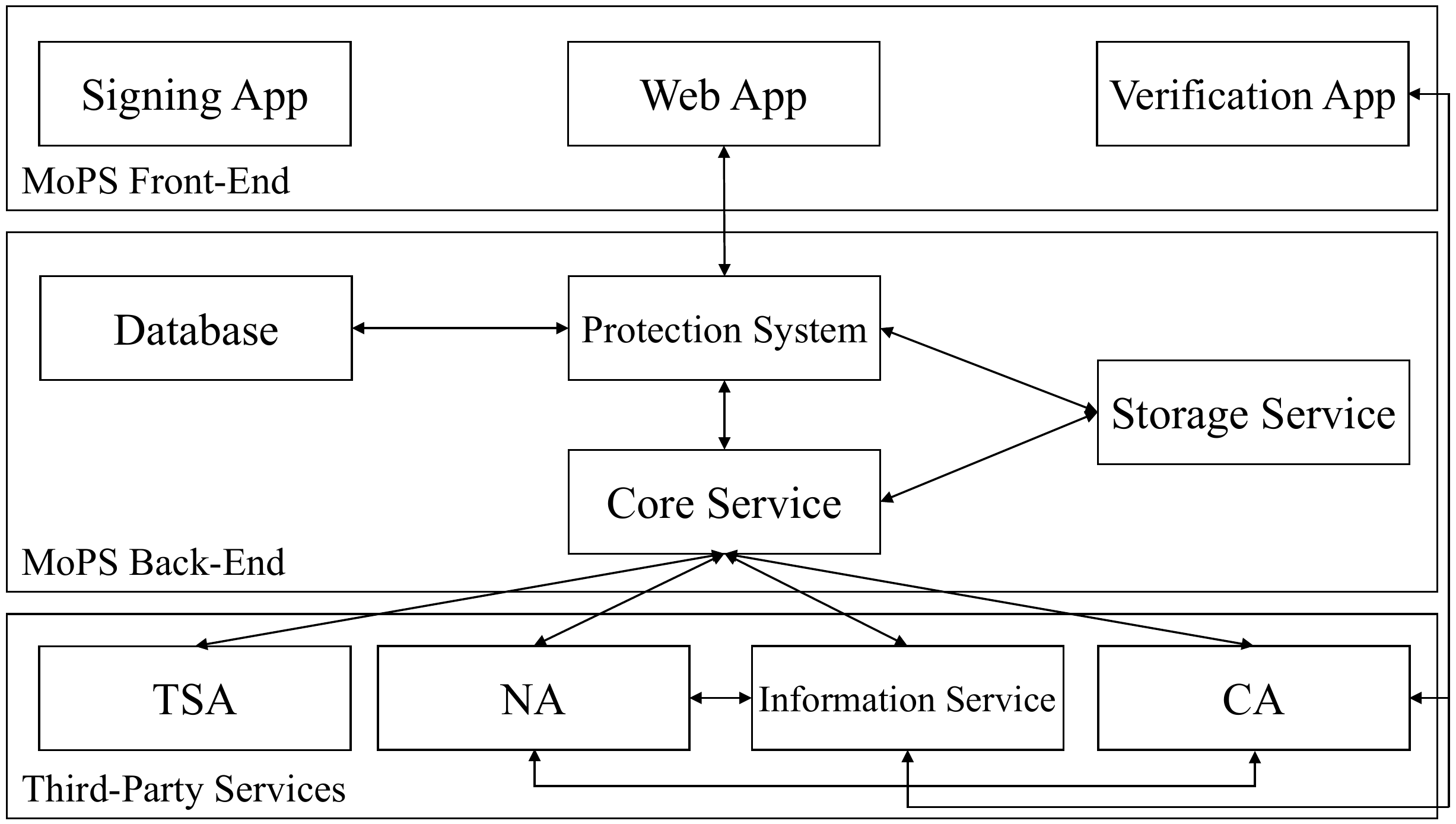}

\caption[MoPS System Architecture]{The different components in the MoPS system architecture grouped into front-end, back-end, and third-party services. Interactions between different components are indicated by arrows.}
\label{fig:apps}
\end{figure}

The MoPS implementation consists of the following applications:
First, we provide a platform-independent graphical desktop application called \emph{Signing App} which enables non-expert users to create signatures.
This application is available for all major desktop operating systems.
Second, we provide a web application called \emph{Web App} to which users can upload signed documents.
The Web App supports non-expert users when creating protection schemes and performing updates by employing wizard-based guidance.
It also allows exporting protected documents for verification or transferring them to other Web App instances.
Accessing the actual protection system via a user interface in form of a web front-end results in platform independence, i.e.\ the implementation can be used on any device providing a web browser.
The third application is a platform-independent graphical desktop application called \emph{Verification App} which allows external retrievers to verify exported documents.
As the Signing App, the Verification App is available for all major desktop operating systems.
An overview of the MoPS components is illustrated in Figure \ref{fig:apps}.

\subsubsection{Signing App}
\label{sec:signing_app}
The Signing App is a graphical desktop application which enables non-expert users to create signatures on documents.
To support a wide range of operating systems, the application was implemented in Java 8 using Java Swing.
When launching the Signing App, the interface invites the users to select a PKCS\#12 key store containing their public-private signature key pair and to add the documents to be signed via drag \& drop.
After starting the signing process, a signature is created for each document.

For creating signatures, the Signing App relies on \linebreak \emph{XAdES4j}\footnote{\url{https://github.com/luisgoncalves/xades4j}}, a Java implementation of XAdES \cite{xades} which also allows creating basic XML signatures containing only an identifier for the employed signature method, the hash value of the signed document, the signing time, the signer's certificate, and the value of the signature on the previously mentioned properties.

The Signing App creates a new \emph{MoPS ZIP} file for each signed file.
The MoPS ZIP format is the file format used for transferring signed and protected documents between the Signing App, the Web App, and the Verification App.
Files in this format are ordinary ZIP files with the extension \enquote{.mops.zip}.
The MoPS ZIP file created by the Signing App contains the signed document and its signature.
The reason for combining a document and its signature into a single file with a custom extension is to facilitate accepting only legit document-signature pairs for the Web App.

\subsubsection{Web App}
\label{sec:web_app}
\begin{figure}
\centering

\includegraphics[width = \columnwidth]{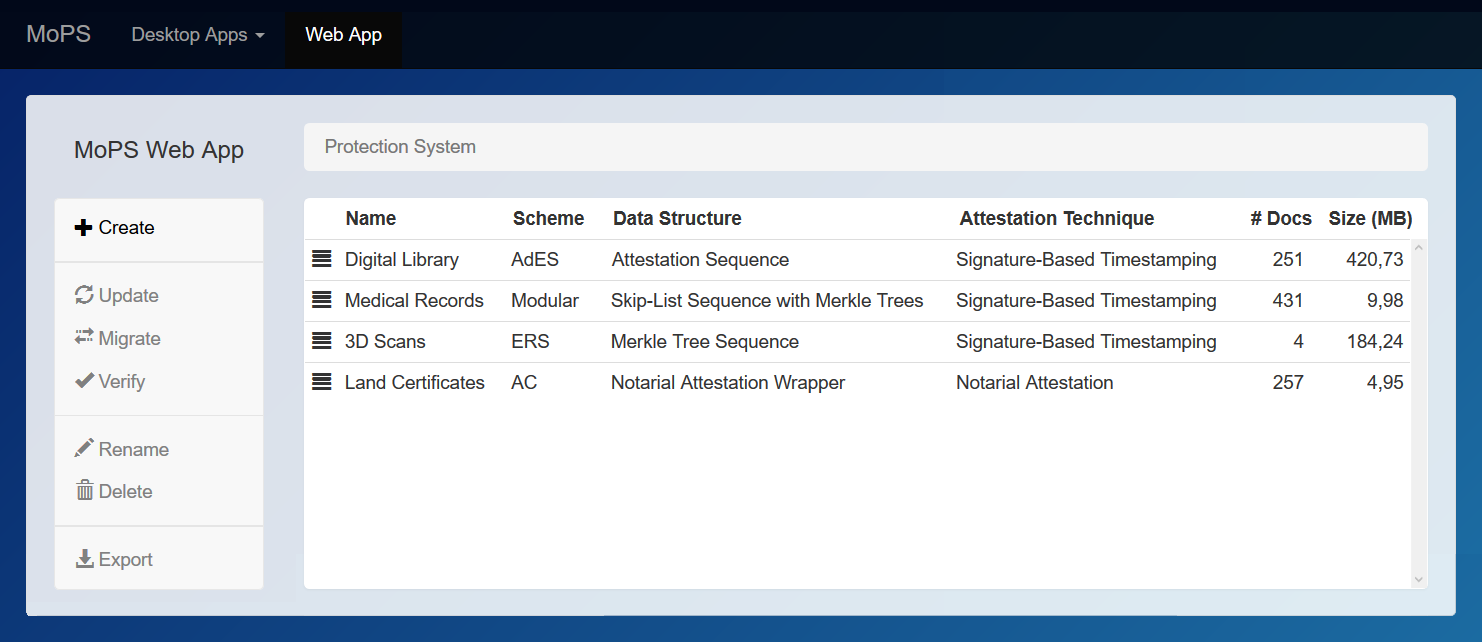}

\caption[MoPS Web App]{A screenshot of the Web App interface showing all currently maintained protection schemes. The side menu allows to manage these protection schemes.}
\label{fig:web_app}
\end{figure}

The Web App is the main front-end component of the MoPS implementation.
It is a wizard-based web interface for the modular protection system.
It allows one to create protection schemes as well as importing, exporting, updating,  migrating, and verifying documents.
The implementation was done using Apache Wicket\footnote{\url{https://wicket.apache.org/}}, a server-side framework for developing interactive web applications using Java 8 and HTML 5.
Figure \ref{fig:web_app} shows a screenshot of the Web App interface.

The protection scheme creation wizard allows non-expert users to create protection schemes with configurations which best suit their needs.
Users can choose between a \emph{simple mode} and an \emph{expert mode}.
Using the simple mode, the users are asked to select the expected access pattern for the documents and the trust assumptions they are willing to make.
Depending on the submitted choices, the system selects the most suitable attestation technique and data structure (or combination of data structures).
The \emph{expert mode} gives expert users the possibility to choose a default configuration for one of the existing schemes surveyed in Section~\ref{sec:modularization}.
Note that due to the lack of support for WVM-based timestamps, the scheme CIS is configured to use signature-based timestamps.
In case the expert users prefer a modular scheme, they can manually select the preferred data structure and attestation technique.
However, impossible combinations are blocked, e.g.\ running $\NAW$ with signature-based timestamps.

In order to import signed documents, the users need to select a) the MoPS ZIP files they want to import, b) the hash function which should be used for hashing the uploaded documents, and c) the remote address of the service issuing the type of attestation they want to use.
After starting the import, first, the selected MoPS ZIP files are uploaded to the Web App.
Then, a proof of existence for the documents is generated by adding them to the selected data structure and creating an attestation with the help of the specified third-party service. 
Finally, the proof of existence is stored in an XML file called \emph{evidence record}.

When exporting one or more protected documents, i.e.\ folders of documents, a download of the data in form of a MoPS ZIP file is offered.
Such a MoPS ZIP file contains all selected documents and their corresponding XML signature files.
In addition, for each folder containing one or more documents, there is an XML file with the extension \enquote{.er.xml} containing the corresponding proof of existence.

The import dialog can also be used to import protected documents which were exported from an arbitrary Web App instance.
In this case, the existing proofs of existence are migrated and updated instead of creating new ones.

The update/migration dialog allows users to update the proof of existence for protected documents.
To do so, they need to decide whether they want to use a new data structure and/or attestation technique and/or hash function and/ \linebreak or attestation service and if so, which one they want to select.

\subsubsection{Verification App}
\label{sec:verification_app}
The Verification App is a graphical desktop application which allows non-expert users to verify protected documents.
This application is necessary to provide external retrievers (i.e.\ users without access to the Web App) the possibility to verify proofs of existence created by the Web App.
When launching the Verification App, the interface prompts the user to open a MoPS ZIP file via drag \& drop.
Also, the user needs to connect to an \emph{information service} (see Section \ref{subsubsec:information_service} for details).
This can be done by selecting either the remote address of an online service or the database shipped with the Verification App which provides recommendations according to Lenstra \cite{lenstra}.
If issues occur during verification, a modal window gives a tabular overview stating why verification failed for which document.

\subsection{Web Services}
\label{sec:web_services}
The MoPS implementation incorporates many web services for performing tasks which require no user interface.
They are RESTful web services implemented in Java 8 using Apache CXF\footnote{\url{https://cxf.apache.org/}}, a framework providing a compliant implementation of the JAX-RS\footnote{\url{https://jsr311.java.net/}} standard.
These web services can be categorized into two groups: third-party services and back-end services.
Third-party services are needed to provide timestamps, notarial attestations, information about the lifetime of cryptographic primitives and parameters, certificates, and revocation information.
The back-end services provide the functionality offered by the Web App to its users:
A so-called \emph{core service} is responsible for performing operations on evidence records.
Furthermore, a dedicated \emph{storage service} manages data objects and, finally, a \emph{protection system} maintains information about protection schemes and protected documents in a database.

\subsubsection{Third-Party Services}
\label{subsec:third_party_services}
The MoPS implementation includes implementations of all required third-party services as there are no existing implementations available (in case of the NA and the information service) or available services are commercial or have limitations regarding free usage (in case of the CA and the TSA service).
In the following we explain only the services for which currently no other implementations exist, i.e.\ the NA and the information service.
Details regarding the CA and TSA service can be found in Appendix \ref{app:tps}.

\paragraph{Information Service}
\label{subsubsec:information_service}
The information service provides information about how long certain cryptographic primitives and parameters are estimated to be secure.
It is also possible to ask whether or not a certain cryptographic method or parameter was considered secure at a given point in time in the past.
Supported cryptographic primitives and parameters are hash functions as well as signature schemes together with recommended key lengths.
In order to fulfill these tasks, the implementation maintains an inventory showing for each of the supported primitives and parameters until when they are expected to be secure.
These dates can be updated if new attacks are discovered.
This type of online service could, for instance, be operated by governmental agencies such as NIST\footnote{\url{https://www.nist.gov/}}.

\paragraph{Notarial Authority (NA)}
\label{subsubsec:notary}
The NA issues and renews notarial attestations, as described in Section \ref{sec:requirements_workflow}, which are transferred in the form of XML documents.
Our NA also supports the migration process proposed in Section \ref{subsec:migration_ac}.
This process requires sending a MoPS ZIP file containing all documents together with their proof of existence to the NA.
The NA is in possession of multiple RSA signature keys with different key lengths.
Which one is used depends on the hash function submitted in the initialization or renewal request (see Table \ref{tab:digest_methods_rsa_key_lengths} in Appendix \ref{subsubsec:ca} for details).

\subsubsection{Storage Service}
\label{subsec:storage_service}
The storage service provides a simple object storage solution:
When uploading a file, it will be stored in a single folder on the service host and a random string will be returned.
This string serves as an identifier for further operations on the object.
In a cloud or enterprise deployment scenario, the storage service could easily be adopted to work as a proxy for Amazon S3\footnote{\url{https://aws.amazon.com/s3/}} or a distributed network file system such as Ceph\footnote{\url{https://www.ceph.com/}}, thereby providing reliability and scalability for extremely large data sets.

\subsubsection{Core Service}
\label{subsec:core_service}
The core service is responsible for performing operations on evidence records: it
a) initializes proofs of existence for documents by creating new data structures according to a given protection scheme configuration,
b) adds documents to an existing data structure,
c) migrates proofs of existence from one protection scheme configuration to another,
d) updates proofs of existence according to the given update parameters,
and
e) verifies a complete proof of existence or only a specific document protected by an evidence record.

For creating attestations, the core service relies on the third-party TSAs and notarial authorities specified in the update parameters.
A storage service is used for hashing objects referenced by proofs of existence and for reading or writing evidence records.
An information service is employed for getting information about cryptographic primitives and parameters during verification.

Following the only formally standardized protection \linebreak schemes AdES and ERS, MoPS stores proofs of existence in the XML format.
The XML schema defining valid XML evidence records can be found online at \url{http://encrypto.de/code/MoPS}.
The Java JDK comes with a binding compiler called \emph{xjc} which was used to generate \emph{Java Architecture for XML Binding} (JAXB) annotated Java classes from the XML schema.
These annotations are then used to automatically create XML representations from Java objects and vice versa.

\subsubsection{Protection System}
\label{subsec:archiving_system}
The protection system is the service back-end for the Web App.
It is responsible for maintaining (i.e.\ creating, renaming, updating, and deleting) information about protection schemes and protected documents in a database.
The protection system is also responsible for importing and exporting protected documents.
For performing operations on evidence records, the protection system uses the core service.

In addition, whenever a document or a folder of documents is imported or updated, the protection system performs validity estimations in order to predict how long a proof of existence is considered valid.
To do so, the latest attestation is extracted from the evidence record.
Then, the protection system sets the estimation value to the earliest of the following dates obtained from an information service:
a) the date until which the hash function used for creating the attestation is considered secure,
b) the date until which the signature scheme used for signing the attestation is considered secure,
c) the date until which the key length of the signature key for signing the attestation is considered secure, and
d) the date after which the certificate for the signature key is no longer valid.

\subsection{Performance Evaluation}
\label{subsec:performance}
In this section, we provide the results of a short performance evaluation.
It is not intended to be a comprehensive comparison between different protection schemes, as this was done in \cite{hpcc2014, cose2015}.
Instead, we want to give an intuition for the additional computation time (given today's hardware and not considering potential increases of computer speed) and storage space a user needs to invest in order to achieve long-term protection using our implementation.

For the evaluation we run each data structure for 100 years using the recommended access pattern for each of them.
All sequence-based data structures use signature-based timestamps, as our implementation does not support WVM-based timestamps.
For the public signature keys, we use certificates with a lifetime of two years.
Thus, signatures must be renewed after two years by the latest.
As to the selection of hash functions and key lengths, we follow the predictions by Lenstra \cite{lenstra} and therefore use SHA-256 with RSA 2048 until 2038 before switching to SHA-384 with RSA 4096 until 2084 before finally switching to SHA-512 with RSA 8192.
Thus, in our evaluation there occur two hash renewals.

For $\AS$ and $\NAW$, we protect and verify a single document.
$\MTS$ protects a set of 100 documents added during initialization while only one of them is verified.
For $\MDS$ and $\SLS$, we add a single document every year.
For $\MDS$, we verify the last document, whereas for $\SLS$ we verify the first one.
Each document is signed and has a size of \SI{1}{\mega\byte}.

The evaluation test suite and all web services were hosted on a single Tomcat 8 application server on Debian 8 powered by an eight core AMD FX\texttrademark{} 8350 CPU @ \SI{4.00}{\giga\hertz} with \SI{16}{\giga\byte} of RAM.
Therefore, the network connection between the different parties had minimal influence on the results.
More realistic measurements in a distributed environment are part of future work.

The results are summarized in Table \ref{tab:performance_evaluation}.
The table shows average values for 10 executions.
Regarding initialization, $\AS$ and $\MDS$ perform equally, as expected.
$\SLS$ is a bit slower, as initializing the parallel hash chains requires additional effort.
$\NAW$ is even slower than $\SLS$, because the NA performs a verification step before creating its attestation.
Of course, $\MTS$ is the slowest data structure regarding initialization, as it creates a Merkle tree for 100 documents, instead of protecting only one initial document.
Regarding verification, $\MTS$ is a bit slower than $\AS$ as for $\MTS$, authentication paths must be used to reconstruct the attested hash values.
Although $\SLS$ contains twice as many attestations as $\AS$ and $\MTS$, the verification of the first document is only 14\% and 12\% slower, respectively.
The reason why $\MDS$ is slower than $\NAW$ is that for $\MDS$ a much larger XML file needs to be opened and parsed before the actual verification procedure can start.

\begin{table}
\centering
\caption{Performance evaluation results. Runtimes are given in ms and sizes in KB.}

\resizebox{\columnwidth}{!}{
\begin{tabular}{@{}lrrrrr@{}}
\toprule
Data Structure					&	$\AS$	&	$\MTS$	&	$\MDS$	&	$\SLS$	&	$\NAW$	\\

\midrule

Initialization					&	\numprint{154}	&	\numprint{2330}	&	  \numprint{154}	&	  \numprint{158}	&	  \numprint{183}	\\
Updates 						&	\numprint{32972}	&	\numprint{40119}	&	\numprint{75652}	&	\numprint{78022}	&	\numprint{32641}	\\
\quad	Attestation renewals	&	\numprint{31122}	&	\numprint{35319}	&	\numprint{72388}	&	\numprint{72980}	&	\numprint{32159}	\\
\quad	Hash renewals			&	\numprint{1850}	&	\numprint{4800}	&	\numprint{3264}	&	\numprint{5042}	&	\numprint{482}	\\
Verification					&	\numprint{1282}	&	\numprint{1315}	&	\numprint{203}	&	\numprint{1467}	&	\numprint{174}	\\
Proof of existence (size)		&	\numprint{680}	&	\numprint{1609}	&	\numprint{2115}	&	\numprint{2127}	&	\numprint{9}	\\

\bottomrule
\end{tabular}
}

\label{tab:performance_evaluation}
\end{table}

\section{Conclusions and Future Work}
\label{sec:conclusion}
In this work we proposed the first modular protection scheme for long-term storage.
More precisely, we provide a set of techniques to build protection schemes which can be plugged together, combined, and migrated.
As a proof of concept, we also implemented MoPS and provide performance measurements.

For future work we plan to integrate techniques that provide long-term confidentiality protection and to further improve our implementation, e.g.\ by supporting automated renewals of proofs.

\newpage

\section*{Acknowledgements}
This work has been co-funded by the DFG as part of projects S5 and S6 within the CRC 1119 CROSSING, by the European Union's Horizon 2020 research and innovation program under Grant Agreement No 644962, and by the German Federal Ministry of Education and Research (BMBF) as well as by the Hessen State Ministry for Higher Education, Research and the Arts (HMWK) within CRISP.

\bibliographystyle{abbrv}

\newpage

\begin{appendix}

\section{Additional Third-Party \\ Services}
\label{app:tps}
Our MoPS prototype includes implementations for all required third-party services.
In this section we provide details regarding the CA and the TSA service.

\subsection{Certification Authority (CA)}
\label{subsubsec:ca}
The main purpose of the CA service in this context is to offer downloads of X.509\footnote{\url{https://tools.ietf.org/html/rfc5280}} certificates and CRLs.
The download URL of the former is embedded in the \emph{authority information access} (AIA) extension and of latter in the \emph{CRL distri\-bution point} (CDP) extension.
Downloading CA certificates and CRLs is necessary for parties who need to collect verification data for certificates.

In addition, the CA prototype implementation fulfills other tasks which cannot be triggered via the service interface: it issues certificates for services and users, revokes certificates (e.g.\ after a key compromise), and updates CRLs.
The certificates and CRLs are created and maintained using the cryptography and SSL/TLS toolkit OpenSSL\footnote{\url{https://www.openssl.org/}}.

Whereas in a real world deployment each CA hosts its own dedicated service instance, our prototype provides certificate and CRL downloads using a single service instance for three hierarchical ordered CAs to simulate the existence of a real PKI.

In fact, our implementation operates multiple PKIs with the same entities and in the same hierarchy.
Each of them uses different hash functions and RSA signatures with different signature key lengths.
On the one hand, this allows one to use the implementation for simulating the aging of cryptography.
On the other hand, end users can choose from different cryptographic primitives and parameters depending on the security level they feel most comfortable with.
For example, a user might want to use SHA-512 with RSA 8192 to sign an important document although SHA-256 with RSA-2048 would be an appropriate choice according to all current recommendations\footnote{\url{https://www.keylength.com/}}.
Table \ref{tab:digest_methods_rsa_key_lengths} shows the supported hash functions and correspondingly used RSA key lengths.
Note that creating a signature using cryptographic primitives with parameters which are expected to be still secure for a very long time does not inevitably imply that less updates are necessary to prolong the protection provided by this signature:
the lifetime of the corresponding certificates is still a limiting factor.

\vfill\eject

\begin{table}[h]
\centering
\caption{The supported hash functions and correspondingly used RSA key lengths.}
\begin{tabular}{@{}lr@{}}

\toprule
Hash function	&	RSA key length (in bit)	\\

\midrule
SHA-256			&	2048					\\
SHA-386			&	4096					\\
SHA-512			&	8192					\\

\bottomrule
\end{tabular}

\label{tab:digest_methods_rsa_key_lengths}
\end{table}

\subsection{Time-Stamping Authority (TSA)}
\label{subsubsec:tsa}
The TSA service provides signature-based timestamps.
It behaves according to the \emph{Time-Stamp Protocol} (TSP\footnote{\url{https://www.ietf.org/rfc/rfc3161.txt}}).
That is, it receives binary encoded attestation requests and returns binary encoded attestations in form of signature-based timestamps.
A timestamp request contains a hash to be attested and the attestation returned contains a so-called timestamp token which in turn contains the timestamped hash, the time at which the timestamp was generated, the TSA's signature, and the TSA's certificate.
Just as the NA, the TSA service is in possession of multiple RSA signature keys with different key lengths.

\end{appendix} 

\end{document}